\documentclass{emulateapj}
\usepackage{graphicx}
\usepackage{apjfonts}
\usepackage{amssymb}
\usepackage{natbib}
\footnotesize
\newdimen\minuswidth    %define @ width of minus sign for tables
\setbox0=\hbox{$-$}
\minuswidth=\wd0
\catcode`@=\active
\def@{\kern\minuswidth}
\newdimen\digitwidth    %define ! a one digit width for tables
\setbox0=\hbox{\rm0}
\digitwidth=\wd0
\catcode`!=\active
\def!{\kern\digitwidth}
\normalsize
\begin{document}
\author{P. A. Gentile\altaffilmark{1}, M. S. E. Roberts\altaffilmark{2,3}, M. A. McLaughlin\altaffilmark{1}, F. Camilo\altaffilmark{4,5}, J. W. T. Hessels\altaffilmark{6,7}, M. Kerr\altaffilmark{8}, S. M. Ransom\altaffilmark{9}, P. S. Ray\altaffilmark{10}, I. H. Stairs\altaffilmark{11}}

\altaffiltext{1}{Dept. of Physics, West Virginia University, Morgantown, WV 26506, USA} % Me and Maura
\altaffiltext{2}{Eureka Scientific Inc., 2452 Delmer Street, Suite 100, Oakland, CA 94602-3017, USA}
\altaffiltext{3}{New York University Abu Dhabi, P.O. Box 129188, Abu Dhabi, United Arab Emirates} % M. Roberts
\altaffiltext{4}{Columbia Astrophysics Laboratory, Columbia University, New York, NY 10027, USA} % F. Camilo
\altaffiltext{5}{Arecibo Observatory, HC3 Box 53995, Arecibo, PR 00612, USA} % F. Camilo
\altaffiltext{6}{ASTRON, the Netherlands Institute for Radio Astronomy, Postbus 2, 7990 AA, Dwingeloo, The Netherlands}% J. Hessels
\altaffiltext{7}{Astronomical Institute ``Anton Pannekoek," University of Amsterdam, Science Park 904, 1098 XH Amsterdam, The Netherlands} % J. Hessels
\altaffiltext{8}{Kavli Institute for Particle Astrophysics and Cosmology, Stanford University, Stanford, CA 94305, USA} % M. Kerr
\altaffiltext{9}{National Radio Astronomy Observatory, Charlottesville, VA 22903, USA} % S. Ransom
\altaffiltext{10}{Space Science Division, Naval Research Laboratory, Washington, DC 20375-5352, USA} % P. Ray
\altaffiltext{11}{Department of Physics and Astronomy, University of British Columbia, Vancouver, BC, V6T 1Z1, Canada} % I. Stairs

\let\leqslant=\leq %Authors with AMS fonts and mssymb.tex can comment
%                 out this line to get the correct symbol for Monthly
%                 Notices.
\date{}

\title[Chandra X-Ray Observations of Black Widow Pulsars]
{X-Ray Observations of Black Widow Pulsars}
\newcommand{\setthebls}{
%                 de-comment this line for double spacing:
%\baselineskip=20pt
}
\setthebls

\begin{abstract}

We describe the first X-ray observations of five short orbital period ($P_B < 1$~day), $\gamma$-ray emitting, binary millisecond pulsars. Four of these - PSRs~J0023+0923, J1124$-$3653, J1810+1744, and J2256$-$1024 - are `black-widow' pulsars, with degenerate companions of mass $\ll0.1~M_{\odot}$, three of which exhibit radio eclipses. The fifth source, PSR~J2215+5135, is an eclipsing `redback'  with a near Roche-lobe filling $\sim$0.2 solar mass non-degenerate companion.  Data were taken using the \textit{Chandra X-Ray Observatory} and covered a full binary orbit for each pulsar. Two pulsars, PSRs~J2215+5135 and J2256$-$1024, show significant orbital variability while PSR~J1124$-$3653 shows marginal orbital variability. The lightcurves for these three pulsars have X-ray flux minima coinciding with the phases of the radio eclipses. This phenomenon is consistent with an intrabinary shock emission interpretation for the X-rays. The other two pulsars, PSRs J0023+0923 and J1810+1744, are fainter and do not demonstrate variability at a level we can detect in these data. All five spectra are fit with three separate models: a power-law model, a blackbody model, and a combined model with both power-law and blackbody components. The preferred spectral fits yield power-law indices that range from 1.3 to 3.2 and blackbody temperatures in the hundreds of eV. The spectrum for PSR~J2215+5135 shows a significant hard X-ray component, with a large number of counts above 2 keV, which is additional evidence for the presence of intrabinary shock emission. This is similar to what has been detected in the low-mass X-ray binary to millisecond pulsar transition object PSR J1023+0038.
\end{abstract}


\section{Introduction}

Of the roughly 2000 radio pulsars known, about 10\% are millisecond pulsars (MSPs) \citep{mhth05}\footnote{http://www.atnf.csiro.au/research/pulsar/psrcat}, old neutron stars which have been spun-up, or `recycled', through accretion of material from a companion \citep{acrs82}. Many details of this recycling process remain unknown, but it is clear that most known MSPs have degenerate white dwarf companions with masses between 0.2 and 1~$M_\odot$. However, $\sim$1/6 of the known MSPs in the Galactic field are isolated\footnote{See http://astro.phys.wvu.edu/GalacticMSPs/}. The process through which these MSPs were formed is unclear. One potentially important method is the ablation of the pulsar companion after the end of the recycling process by energetic particles and/or $\gamma$-rays produced in the pulsar magnetosphere \citep{rst89a}.

The identification of MSPs as strong $\gamma$-ray sources \citep{aaa+09b, khv+00} motivates searches for radio pulsations in unidentified {\it Fermi} sources with spectral and temporal properties matching those of known $\gamma$-ray MSPs.  Bangale et al. (2013)\nocite{bcd+13} observed 49 sources at 350 MHz with the Green Bank Telescope (GBT) and detected 17 MSPs, 10 of which were new discoveries and 16 of which are in binary systems, with seven of them having short orbital periods (P$_{b} < 1$ day). Three of these pulsars (PSRs~J0023+0923, J1124$-$3653, J1810+1744) and one (PSR~J2256$-$1024) found in a 350-MHz GBT drift-scan survey (\cite{blr+13}; Stairs et al., in prep) and re-detected in the Bangale et al. survey  have very small companion masses (M$^{c}_{min} \ll 0.1 M_{\odot}$) and three have pronounced radio eclipses, classifying them as black-widow pulsars \citep{sgk+03}. One other pulsar discovered in this survey (PSR~J2215+5135) has a short orbital period and eclipses, but a larger companion mass (M$^{c}_{min} = 0.208 M_{\odot}$) (Table 1). Optical observations of the companion suggest it is non-degenerate and nearly Roche-lobe filling and hence may be in an only temporary non-accreting, radio-emitting phase \citep{bvr+13}. 

 The first pulsar showing evidence for the ablation process was the original black-widow pulsar PSR~B1957+20, which shows radio eclipses due to absorption in the wind of the companion and dramatic pulse delays around the time of eclipse due to propagation through the wind \citep{fbb+90}. {\it XMM-Newton} \citep{hb07} and {\it Chandra} observations \citep{sgk+03,hkt+12} revealed unresolved synchrotron emission that is variable throughout the orbit. 
%% with roughly half of the emission due to the interaction of the pulsar and stellar winds and half from the pulsar itself. 
On average, the orbital modulation is broadly sinusoidal,  peaking near  superior conjunction when the companion is between the pulsar and observer, but with a narrow dip over $\sim$0.1 of the orbit at superior conjunction. This emission is interpreted as coming from an intrabinary shock of the pulsar's wind close to the nearly Roche-lobe filling companion's surface \citep{vbk11}. In addition, the {\it Chandra} observations resolved an extended tail of X-ray emission arising from the pulsar outflow shocking the interstellar medium, the first demonstration that millisecond pulsars can produce pulsar wind nebulae.  Furthermore, magnetospheric pulsations in $\gamma$-rays and X-rays have been detected from the point source \citep{gjv+12}.

%% A possible explanation for the smaller dip width in the lightcurve of PSR~B1957+20 relative %%to PSR~J1023+0038 (the dip for PSR~B1957+20 was $\sim0.1$ of an orbit while the dip in %%PSR~J1023+0038 was $\sim0.3$ of an orbit) is that PSR~B1957+20 could have a higher %%inclination than PSR~J1023+0038. 

An important link in the MSP formation scenario was made with the discovery of a radio pulsar (PSR~J1023+0038) that showed evidence for having an accretion disk in the recent past \citep{asr+09}. This very fast ($P_{spin} = 1.69 $ ms) eclipsing radio pulsar is in a 4.8-hr orbit around a nearly Roche-lobe filling, non-degenerate companion, and is the prototype of the `redback' class of binary MSPs \citep{rob11}.  {\it XMM-Newton} \citep{akb+10} and {\it Chandra} \citep{bah+11} observations 
%%(OBSID 11075, PI Lorimer) 
of this system revealed significant orbital variability over multiple consecutive orbits, with a pronounced dip in the X-ray flux at superior conjunction (orbital phases of $\sim$0.1 to 0.4), when the pulsar is behind the companion and the intrabinary shock produced through the interaction of stellar outflows is obscured \citep{bah+11}. Because the angular extent of the pulsar as seen from the companion star is small, the width of this dip suggests that the X-ray emission region is much closer to the companion star than to the MSP. This evidence is strengthened further when considering the inclination of the binary system ($i$ $\sim$46$^{\circ}$, constrained through optical radial velocity measurements \citep{asr+09}). The X-ray spectrum consists of a dominant non-thermal component from the shock and at least one thermal component, likely originating from the heated pulsar polar caps. X-ray pulsations were also observed in the {\it XMM-Newton} data, indicating that some of the non-thermal point source emission is magnetospheric. For this source, no evidence for extended X-ray emission has been seen in the {\it Chandra} data \citep{bah+11}.

%GENERAL: ARE THERE ANY ESTIMATES OF THE INCLINATION OF 1957? YES, BUT THE GREATER DISTANCE AND SMALLER COMPANION MASS ARE THE REASONS FOR SMALLER ECLIPSE.

In general, the shock X-ray luminosity for a binary pulsar system will depend on the fraction of the wind intercepted by the companion, the spin-down energy loss rate ($\dot{E}$) of the pulsar, and both the post-shock magnetic field strength and the ratio of electromagnetic flux to kinetic energy flux, $\sigma$ \citep{at93,kc84a}. For PSR~B1957+20, measurements of the X-ray orbital variability show that the efficiency of X-ray production at the shock is similar to that of pulsar wind nebulae around young pulsars, but it is not clear if this is true in all cases.

The body of knowledge regarding black-widow pulsars is still lacking.  It remains to be shown, for example, whether intrabinary shocks can produce significant mass loss only from companions which are nearly filling their Roche-lobe. 
%%arise in  can best be modeled as being close to the surface of the companion star or rather as %%being close to the inner Lagrangian point. 
It also is not clear whether or not the winds from these pulsars are dominated by kinetic or magnetic energy. 

Until very recently, studies were limited by the rarity of these systems. In the last few years however, many nearby systems have been discovered, more than tripling the known population \citep{rap+12}.
In Section ~\ref{sec:observations}, we summarize the observations and analysis procedures. In Section ~\ref{sec:results}, we present the results of the spectral and light curve analyses. In Section ~\ref{sec:conclusions}, we offer conclusions. For each of these sources, we compare the X-ray properties to those of PSR~B1957+20 and PSR~J1023+0038, currently the two best-studied systems.

\section{Observations and Analysis}\label{sec:observations}

We observed PSRs~J0023+0923 (OBSID 14363), J1124$-$3653 (OBSID 13722), J1810+1744 (OBSID 12465), J2215+5135 (OBSID 12466), and J2256$-$1024 (OBSID 12467) for 15~ks, 22~ks, 22~ks, 19~ks, and 22~ks respectively, i.e. at least a full orbit in each case. The data were taken using \textit{Chandra}'s ACIS-S detector and analyzed using \textit{Chandra}'s data analysis suite, CIAO (version 4.2). Source regions were selected by first determining the locations of each source using radio timing positions. Background regions were chosen so that the regions were located on the same chip as the source and did not enclose any point sources.
Once source and background regions were selected, point-spread functions (PSFs) were then created for each source using CIAO's \verb mkpsf  function. Since the CIAO PSF library has PSFs for five discrete energies, we evaluated the PSF at energies which approximately correspond to where the count rate is a maximum. The energies chosen for PSRs~J0023+0923, J1124$-$3653, J1810+1744, J2215+5135, and J2256$-$1024 were 0.75, 1.25, 1.5, 0.8, and 1 keV, respectively. Once energies were chosen, the \verb mkpsf  tool linearly interpolates a PSF from the five PSFs in the library. We then used these to compare the intensity of the source and PSF in two orthogonal directions. 
%TOMALLORY: DO YOU THINK WE SHOULD INCLUDE THE PSF PLOTS?

%\begin{figure*}
%\centering
%\includegraphics[width=4.1cm]{J0023+09scrunchedcombined.ps}
%\includegraphics[width=4.1cm]{J1810+17scrunchedcombined.ps}
%\includegraphics[width=4.1cm]{J2215+51scrunchedcombined.ps}
%\includegraphics[width=4.1cm]{J2256-10scrunchedcombined.ps}
%\caption{Comparison of PSFs and source images for the X and Y axes of PSRs~J0023+0923, J1810+1744, J2215+5135, and J2256$-$1024. The PSF of each source (solid line) for each axis was compared to the intensity of the source summed along each axis (dashed line) using the KS test.}
%\label{fig:figure1}
%\end{figure*}

\begin{table*}
\begin{center}
\footnotesize
\begin{tabular}{ccccccccccc}
\hline
\hline
\bf Name & \bf P$_{spin}$ & \bf log$_{\bf 10} \dot{\bf E}$ & \bf DM  & nH & \bf D &   \bf P$_{orb}$ &  \bf M$^{c}_{min}$ & \bf T$_{obs}$ & MJD$_{obs}$ & \bf Cts\\
           PSR & (ms) & (erg s$^{-1}$) & (pc cm$^{-3}$) & $10^{20}$ cm$^{-2}$ & (kpc) & (hr) & (M$_{\odot}$) & (ks) &  & \\
\hline

J0023+0923   & 3.05 & 34.2 & 14.3 &  4.4 & 0.7 & 3.3 & 0.016 & 15 & 55893 & 43 \\
J1124$-$3653 & 2.41 & 33.6 & 44.9 & 15.7 & 1.7 & 5.5 & 0.027 & 22 & 56118 & 138 \\
J1810+1744   & 1.66 & 34.6 & 39.7 & 12.2 & 1.9 & 3.6 & 0.035 & 22 & 55740 & 55\\
J2215+5135   & 2.61 & 34.7 & 69.2 & 21.4 & 3.0 & 4.2 & 0.22  & 19 & 55697 & 133\\
J2256$-$1024 & 2.29 & 34.6 & 13.8 &  4.3 & 0.6 & 5.1 & 0.030 & 22 & 55788 & 141\\
\hline
{\it J1023+0038} & {\it 1.68} & {\it 34.6} & {\it 14.3} & 18.0 & {\it 1.3} & {\it 4.8} & {\it 0.2} & {\it 83 } & 55281 & {\it 3270}\\ 
{\it B1957+20} & {\it 1.60} & {\it 35.2} & {\it 29.1} & <1.0 & {\it 2.5}  & {\it 9.1} & {\it 0.020} & {\it 43 } & 52081 & {\it 370}   \\
\hline
\end{tabular}
\end{center}
\label{table:table1}
\tablecomments{Timing and X-ray properties of the five {\it Fermi}-associated radio MSPs, including the pulsar spin period (P$_{spin}$), the logarithm of the spin-down energy loss rate (log$_{10} \rm \dot{E}$), dispersion measure (DM), neutral Hydrogen column density along the line of sight to the source (nH), distance to the pulsar (D), orbital period of the binary system (P$_{orb}$), minimum companion mass (M$^{c}_{min}$), total observation duration (T$_{obs}$), MJD of observation (MJD$_{obs}$), and total background-subtracted counts (Cts). Due to the low number of background-subtracted counts, nH is estimated from DM (see text) and held fixed for each source. PSRs~J1023+0038 and B1957+20 are shown for comparison. Timing properties are from 350-MHz observations with the GBT (see Bangale et al. 2013 and Hessels et al., in prep). Distances are estimated from the DM using the \protect\cite{cl02} model for the Galactic electron density, except for PSR J1023+0038 which is from parallax measurements \citep{dab+12}.}
\end{table*}

\begin{table*}
\begin{center}
\footnotesize
\begin{tabular}{ccccccc}
\multicolumn{7}{c}{Power-law Fit}\\
\hline
\hline
\bf Name  & $\bf \Gamma$ & \bf F$_{\bf x}$ & \bf log$_{\bf 10} \bf L_x$ & $\epsilon$ & $\chi^{2}/$ DOF&\\
PSR &  & ($10^{-14}$ erg s$^{-1}$ cm$^{-2}$) & (erg s$^{-1}$) & ($10^{-5}$) &\\
\hline

J0023+0923   & 3.2$^{+0.6}_{-0.5}$ & $2.2 \pm 0.7$ & $30.1$ & 8   &  1.0 / 7 &\\
J1124$-$3653 & 2.1$^{+0.3}_{-0.3}$ & $4.2 \pm 1.7$ & $30.7$ & 130 & 12.3 / 7 &\\
J1810+1744   & 2.1$^{+0.4}_{-0.4}$ & $2.0 \pm 0.9$ & $31.0$ & 20  &  2.1 / 7 &\\
J2215+5135   & 1.4$^{+0.2}_{-0.2}$ & $9.0 \pm 2.5$ & $32.0$ & 190 &  1.8 / 7 &\\
J2256$-$1024 & 2.7$^{+0.2}_{-0.2}$ & $4.0 \pm 0.4$ & $30.2$ & 4   &  5.1 / 7 &\\

%Old table
%J0023+0923 & -- & 3.1$^{+0.5}_{-0.4}$ & $29.3 \pm 0.1$  & $30.1 \pm 0.1$ &\\
%J1124$-$3653 & --& 1.7$^{+0.2}_{-0.2}$ & $29.7\pm 0.1$ & $31.3 \pm 0.1$ &\\
%J1810+1744 & -- & 1.9$^{+0.4}_{-0.3}$ & $29.2 \pm 0.1$ & $31.0 \pm 0.1$ &\\
%J2215+5135 & -- & 1.0$^{+0.2}_{-0.2}$ & $29.9 \pm 0.1$ & $32.0 \pm 0.1$ &\\
%J2256$-$1024 & -- & 2.8$^{+0.2}_{-0.2}$ & $29.6 \pm 0.05$ & $30.3 \pm 0.1$ &\\
\hline
%{\it J1023+0038} & {\it 1.19$^{+0.03}_{-0.03}$} & {\it 30.7 } & {\it 32.0 } & {\it 100}& {\it -- }\\ 
%{\it B1957+20} & {\it 1.9$^{+0.5}_{-0.5}$} & {\it 29.9 } & {\it 31.8 } & {\it 40}& {\it -- }\\
\hline
\multicolumn{7}{c}{ }\\
\multicolumn{7}{c}{Blackbody Fit}\\
\hline
\hline
\bf Name & \bf kT & \bf F$_{\bf x}$ & \bf log$_{\bf 10} \bf L_x$ & $\epsilon$ & $\chi^{2}/$ DOF&\\
PSR & (eV) & ($10^{-14}$ erg s$^{-1}$ cm$^{-2}$) & (erg s$^{-1}$) & ($10^{-5}$)&\\
\hline

J0023+0923   & 180$^{+60}_{-50}$   & $2.0 \pm 4.5$  & $30.1$ & 8   & \ \ 3.6 / 7 &\\
J1124$-$3653 & 440$^{+100}_{-80}$  & $3.5 \pm 4.8$  & $31.1$ & 300 &    27.5 / 7 &\\
J1810+1744   & 430$^{+130}_{-120}$ & $1.6 \pm 2.1$  & $30.8$ & 20  & \ \ 4.9 / 7 &\\
J2215+5135   & 700$^{+150}_{-130}$ & $7.2 \pm 8.0$  & $31.9$ & 160 &    13.4 / 7 &\\
J2256$-$1024 & 200$^{+20}_{-20}$   & $2.7 \pm 1.7$  & $30.1$ & 3   & \ \ 6.4 / 7 &\\

%Old table:
%J0023+0923 & 190$^{+50}_{-40}$ & -- & $29.1 \pm 0.6$  & $30.0 \pm 0.6$ &\\
%J1124$-$3653 & 490$^{+100}_{-80}$ & -- & $29.5\pm 0.4$ & $31.0 \pm 0.4$ &\\
%J1810+1744 & 400$^{+110}_{-100}$ & -- & $29.1 \pm 0.6$ & $30.8 \pm 0.6$ &\\
%J2215+5135 & 790$^{+150}_{-120}$ & -- & $29.7 \pm 0.4$ & $31.8 \pm 0.4$ &\\
%J2256$-$1024 & 200$^{+20}_{-20}$ & -- & $29.4 \pm 0.3$ & $30.0 \pm 0.3$ &\\
\hline
\end{tabular}
\begin{tabular}{cccccccc}
\multicolumn{8}{c}{ }\\
\multicolumn{8}{c}{Combined Fit}\\
\hline
\hline
\bf Name & $\bf \Gamma$ & \bf F$_{\bf x}$ & \bf log$_{\bf 10} \bf L_x$ & $\epsilon$ & \bf Blackbody Flux & \bf Power-Law Flux & $\chi^{2}/$ DOF\\
PSR &  & ($10^{-14}$ erg s$^{-1}$ cm$^{-2}$) & (erg s$^{-1}$) & ($10^{-5}$) & ($10^{-14}$ erg s$^{-1}$ cm$^{-2}$) & ($10^{-14}$ erg s$^{-1}$ cm$^{-2}$) & \\
\hline

J0023+0923   & 1.5                 & $2.0 \pm 0.7$     & $30.1$ & 8   & 1.5$^{+0.5}_{-0.5}$ &  1.0$^{+0.7}_{-0.7}$&  2.7 / 7\\
J1124$-$3653 & 1.3$^{+0.5}_{-0.4}$ & $6.5 \pm 3.1$     & $31.3$ & 500 & 2.3$^{+1.0}_{-1.0}$ &  5.3$^{+5.0}_{-2.6}$&  9.3 / 6\\
J1810+1744   & 1.5                 & $2.0 \pm 0.6$     & $30.9$ & 20  & 0.7$^{+0.4}_{-0.4}$ &  2.0$^{+0.6}_{-0.6}$&  2.4 / 7\\
J2215+5135   & 1.2$^{+0.4}_{-0.3}$ & $8.4 \pm 4.6$     & $32.0$ & 200 & 1.2$^{+1.1}_{-0.8}$ &  9.4$^{+1.5}_{-1.5}$&  1.5 / 6\\
J2256$-$1024 & 1.8$^{+0.7}_{-0.6}$ & $4.4 \pm 2.2$     & $30.3$ & 10  & 2.5$^{+1.0}_{-1.0}$ &  3.3$^{+2.6}_{-1.8}$&  2.0 / 6 \\

%Old table
%J0023+0923 & 230$^{+300}$ & 3.2$^{+8.3}_{-1.9}$ & $29.4 \pm 0.4$  & $30.2 \pm 0.4$ &\\
%J1124$-$3653 & 110$^{+70}_{-60}$ & 1.5$^{+0.3}_{-0.3}$ & $29.8\pm 0.3$ & $31.4 \pm 0.3$ &\\
%J1810+1744 & 330$^{+1080}$ & 1.8$^{+1.6}_{-1.8}$ & $33.6 \pm 0.1$ & $35.3 \pm 0.1$ &\\
%J2215+5135 & 400$^{+2680}$ & 0.7$^{+0.7}_{-0.7}$ & $30.3 \pm 1.0$ & $32.4 \pm 1.0$ &\\
%J2256$-$1024 & 170$^{+20}_{-20}$ & 1.3$^{+1.1}_{-1.3}$ & $30.2 \pm 0.2$ & $30.8 \pm 0.2$ &\\
\hline
\end{tabular}
\end{center}
\label{table:table2}
\tablecomments{Spectral properties of the five {\it Fermi}-associated radio MSPs, including the temperature (kT), power-law index ($\Gamma$), the logarithm of the measured absorbed flux (log$_{10} \rm F_x$), the logarithm of the 0.3--8 keV luminosity (log$_{10} \rm L_x$), the 0.3--8 keV efficiency ($\epsilon$), and the ratio of the $\chi^{2}$ value to the degrees of freedom (DOF) for each fit. The very low $\chi^{2}$ values obtained suggest the fits to be overdetermined. Also included for the combined fit are the contributions to the unabsorbed flux from each component. All fits were performed using \textit{Chandra}'s fitting package, Sherpa. All five sources were fitted with three separate models: a power-law model, a blackbody model, and a combined model with both power-law and blackbody components. The results of all three fits are shown. For the combined fits, values without errors were held constant, as was the temperature for each source (150 eV).}
\end{table*}

Lightcurves were then determined for each source using counts in the 0.3 to 8~keV range, as \textit{Chandra} has very little effective area outside of that range. The number of background-subtracted counts detected for each source ranged from 43 to 141 (Table~\ref{table:table1}). Each lightcurve was binned such that each bin represents one tenth of the observation, so that all bins have equal exposure. For these lightcurves, an orbital phase of 0.25 corresponds to the superior conjunction of the system. These lightcurves were then compared to uniform distributions using the $\chi^{2}$ test and  Kolmogorov-Smirnov (K-S) test \citep{pftv89} to determine their orbital variability. We have left these lightcurves unfolded to show the consistency of the shape from orbit-to-orbit.

Spectra were then analyzed using \textit{Chandra}'s spectral fitting platform, Sherpa. The data were binned with 5 bins between 0.3 and 2 keV and 4 bins between 2 and 8 keV. This binning scheme was used in order to differentiate thermal emission (which we expect below 2 keV) and non-thermal emission (which we expect above 2 keV), which requires multiple bins above 2 keV. The data were then fitted over energies between 0.3 and 8 keV. Due to the small number of background-subtracted counts, we fixed nH, the neutral Hydrogen column density along the line of sight to the source at a constant value set by the dispersion measure, assuming 10 free electrons per neutral Hydrogen atom as is motivated by \cite{hnk13}. The resulting column densities %in %units of $10^{20}$ cm$^{-2}$ were 4.4, 15.7, 12.2, 21.4, and 4.3 for PSRs~J0023+0923, %J1124$-$3653, J1810+1744, and J2256$-$1024, 
are listed in Table~\ref{table:table1}. %Wording is really clumsy there.
 Comparing the values to the total Galactic nH as estimated using the {\it HEASARC} nH tool (based on the maps of \cite{kbh+05,dl90}), we found that this gives reasonable values. We fit each source with three separate models: a power-law model, a blackbody model, and a combined model with both power-law and blackbody components. Due to the low number of background-subtracted counts for PSRs~J0023+0923 and J1810+1744, temperature and $\Gamma$ were held constant at 150 eV and 1.5, respectively, for their combined fits. These values are consistent with temperatures typically seen for millisecond pulsars \citep{bog08,brg07,zc03} and power-law indices typically seen for non-thermal neutron star emission \cite{bgv05,bah+11}.\footnote{Also see http://astro.phys.wvu.edu/XrayMSPs for a list of millisecond pulsars observed in X-Rays and their fit parameters.} Since PSRs~J1124$-$3653, J2215+5135, and J2256$-$1024 all had higher count rates, we kept their temperatures fixed at 150 eV, but let $\Gamma$ vary. Also due to the low number of counts, all fits were done using cstat, which is Sherpa's equivalent to XSPEC's Cash statistic.
%Single component spectral models were compared using the F-test, which is a model %comparison test to determine which of two competing models best characterizes a data set %\citep{pftv89}. We do not have enough photons to determine whether the combined model is %preferred over either of the single-component models for any of the sources. The resultant fit %parameters are reported in Table~\ref{table:table2}. \\

\section{Results}\label{sec:results}
\subsection*{PSR~J0023+0923}\label{sub:J0023+0923}

\begin{figure}
\centering
\includegraphics[scale=.4]{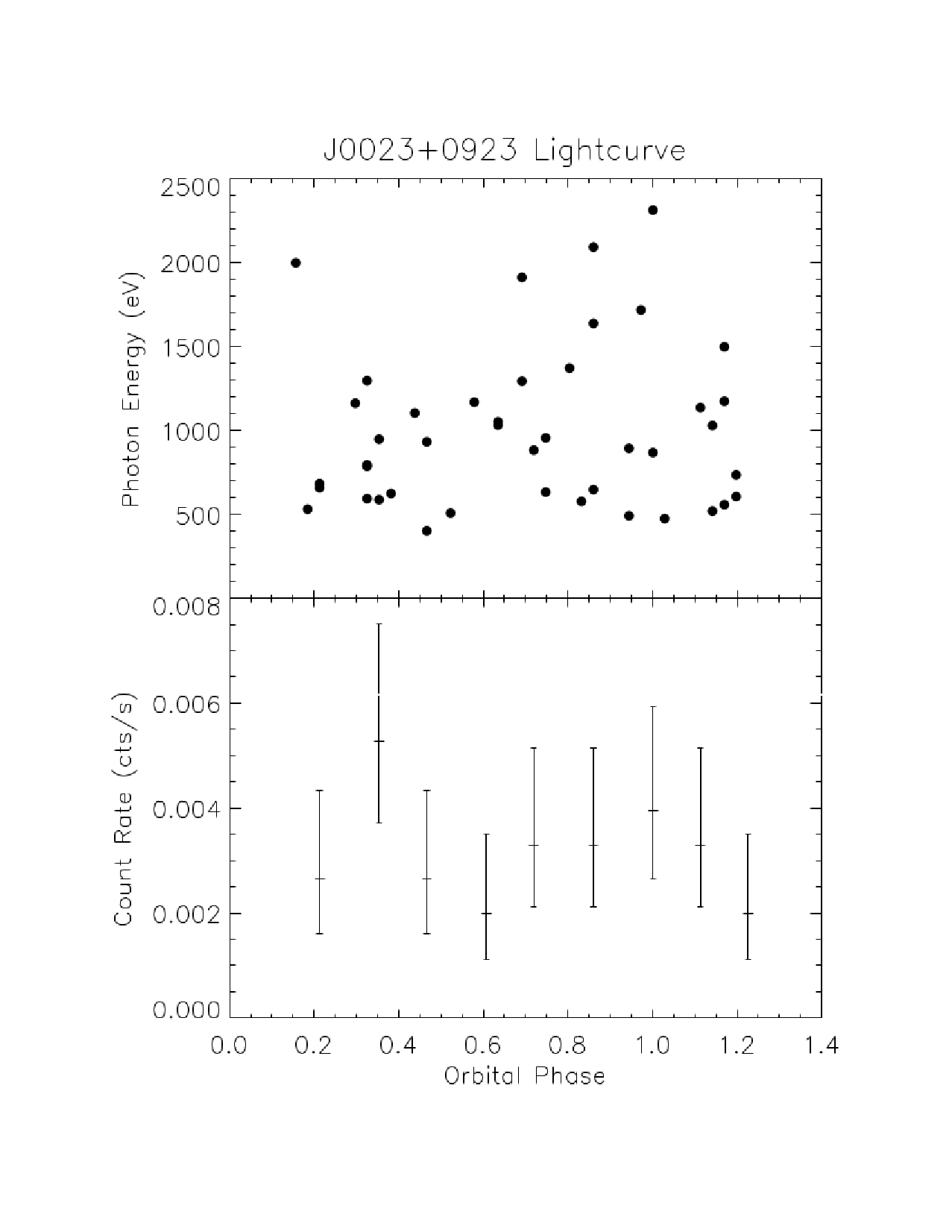}
\includegraphics[scale=.4]{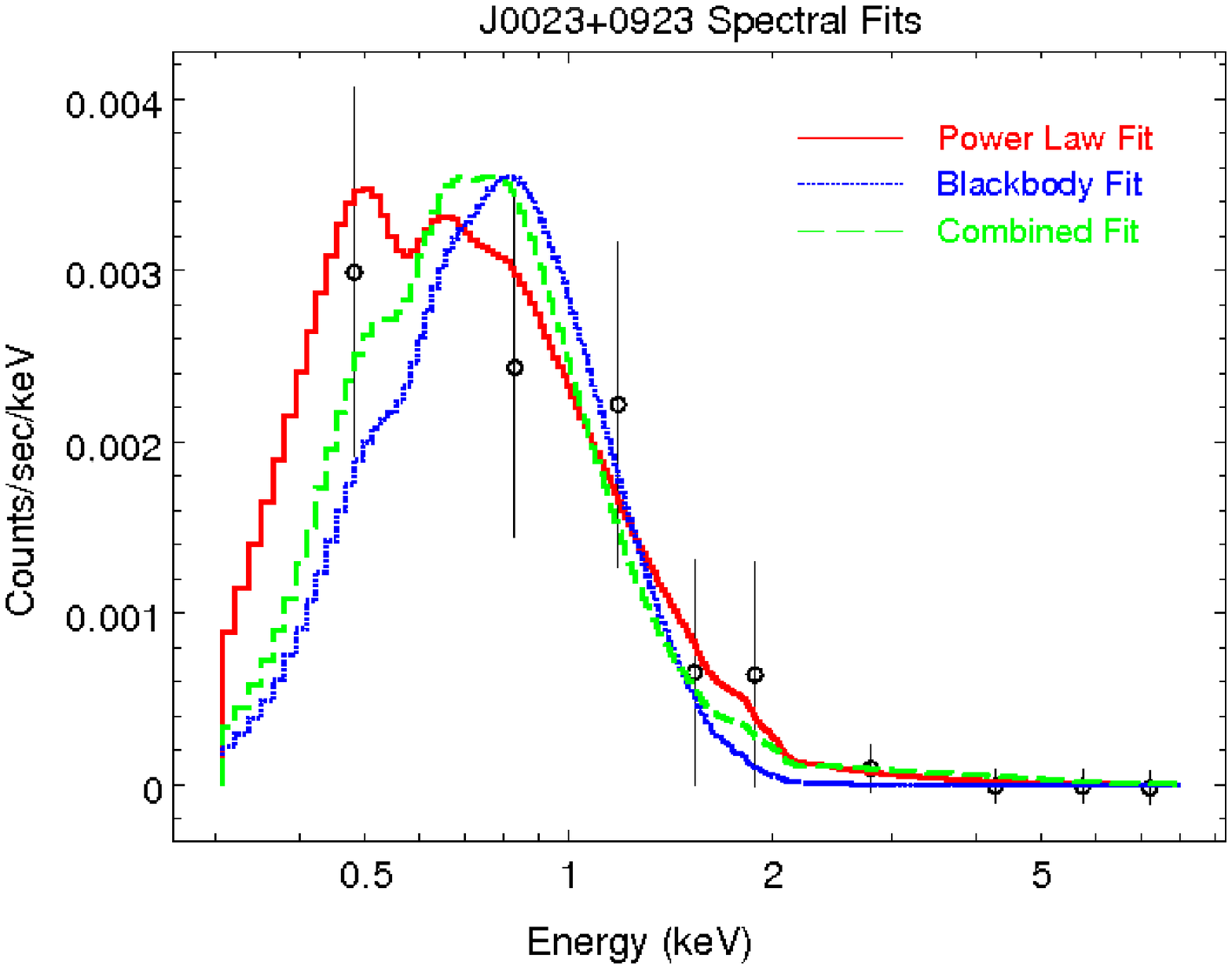}
\caption{Count rate versus orbital phase and spectral fits for PSR~J0023+0923.
% We detected 43 counts and see no significant orbital variation. The K-S test gives a probability of 0.99 of the lightcurve being drawn from a uniform distribution. We fit the spectrum with a blackbody model, a power-law model, and a combined model with power-law and blackbody components. The F-test shows that the combined fit is the preferred fit. {\it Non-Poissonian errors were applied to bins with less than 10 counts in the bin.??}
%GENERAL: ARE THEY REALLY NON-POISSON? I think we are just using the correct non root N errors for small numbers.
%
%The errors seem to have come from this website: http://www-cdf.fnal.gov/physics/statistics/notes/pois_eb.txt. From the looks of things on that site, the formulae that describe the upper and lower error limits were just made up by a committee that thought these errors worked best if you didn't want to use sqrt(n). It doesn't seem like there was any sort of mathematical derivation going on (like you can derive sqrt(n) from Poisson's distribution), so there's an argument for calling these errors "non-Poissonian" errors. Then again, these errors were formulated specifically for Poissonian data, so I think there's an arguemnt either way. What do you think? -Pete
Radio eclipse start and stop times are shown as dotted lines (when applicable). All error bars correspond to 1$\sigma$ errors. 
%DO YOU MEAN 68% CONFIDENCE REGION, AS ESTIMATED FROM THE ERROR TOOL, I.E FROM A DELTA CHI-AQ OF 1 (NOT 2.706 WHICH GIVES THE 90% CONFIDENCE REGION AND IS WHAT IS USUALLY REPORTED IN X-RAY PAPERS??) IF UNEQUAL ERROR BARS,  THEN NOT REALLY 1 SIGMA.
Fit parameters are summarized in Table~\ref{table:table2}.
\label{fig:fig2}}
\end{figure}

The lightcurve appears uniform (within 1$\sigma$ errors), and, according to the K-S test, has a probability of 0.99 of being drawn from a uniform distribution. This is consistent with this pulsar showing no radio eclipse, even at the relatively low observing frequency of 350MHz. A two-dimensional K-S test yields the probability of being a point source of 0.99 in the x-direction and 0.31 in the y-direction. Therefore, we conclude that there is no evidence for extended emission. There is no detected emission above 2.5 keV, and so we effectively had only 6 bins (4 DOF) with which to fit, the very low $\chi^{2}$ values obtained suggest the fits to be overdetermined. While formally the power-law fit is slightly better than the blackbody fit, the lack of high energy counts and the steep power-law index ($\Gamma$ $\sim$3) of the power-law fit, and the reasonable temperature obtained from the blackbody fit all suggest the emission is predominantly thermal. 
%We fit PSR~J0023+0923's spectrum with three different models: a power-law model, a %blackbody model, and a model that combines both power-law and blackbody components. Since the count rate was so low, we fixed temperature and $\Gamma$ for the combined fit, which also suggests most of the flux comes from the thermal component.

\subsection*{PSR~J1124$-$3653}\label{sub:J1124-3653}

\begin{figure}
\centering
\includegraphics[scale=.4]{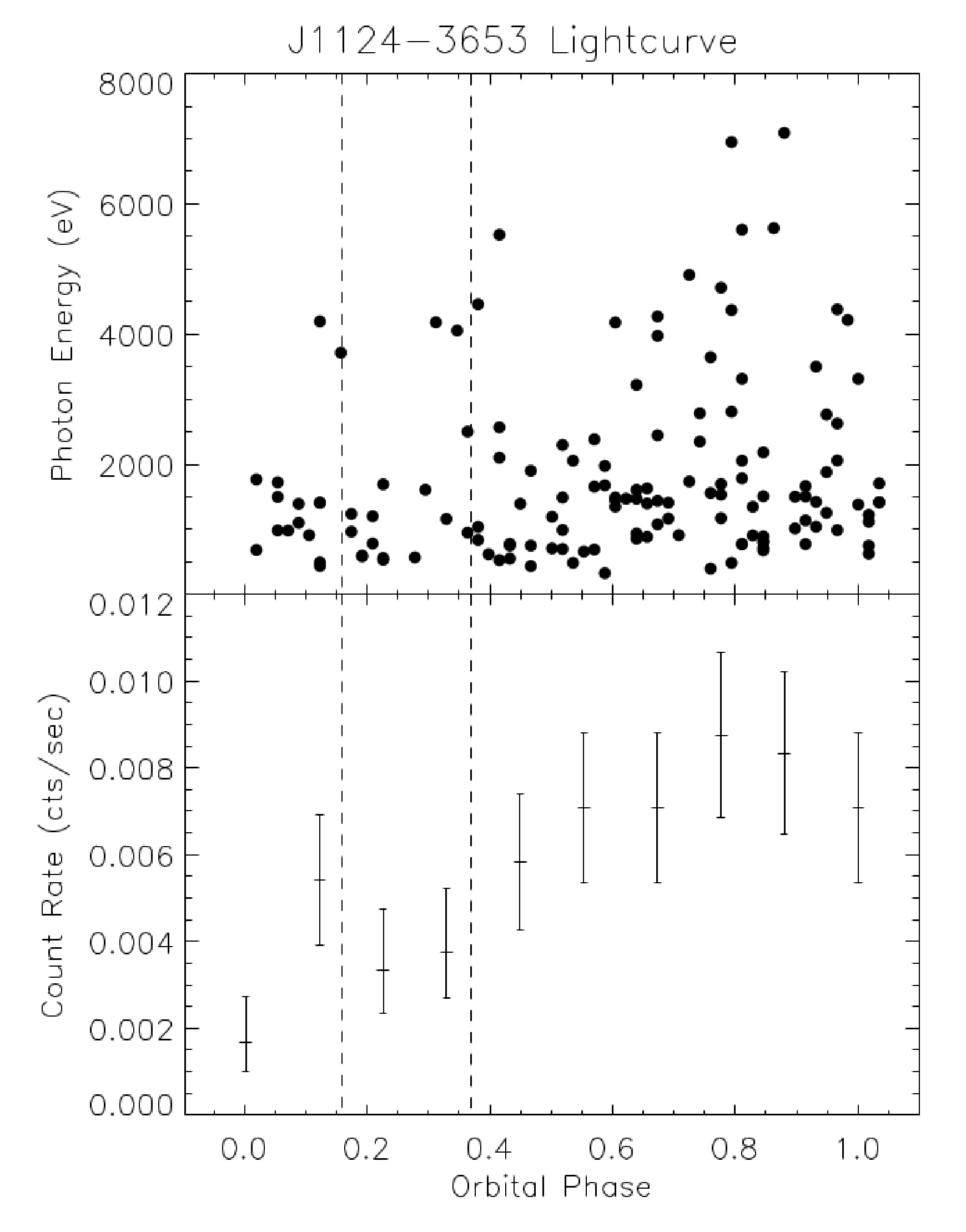}
\includegraphics[scale=.4]{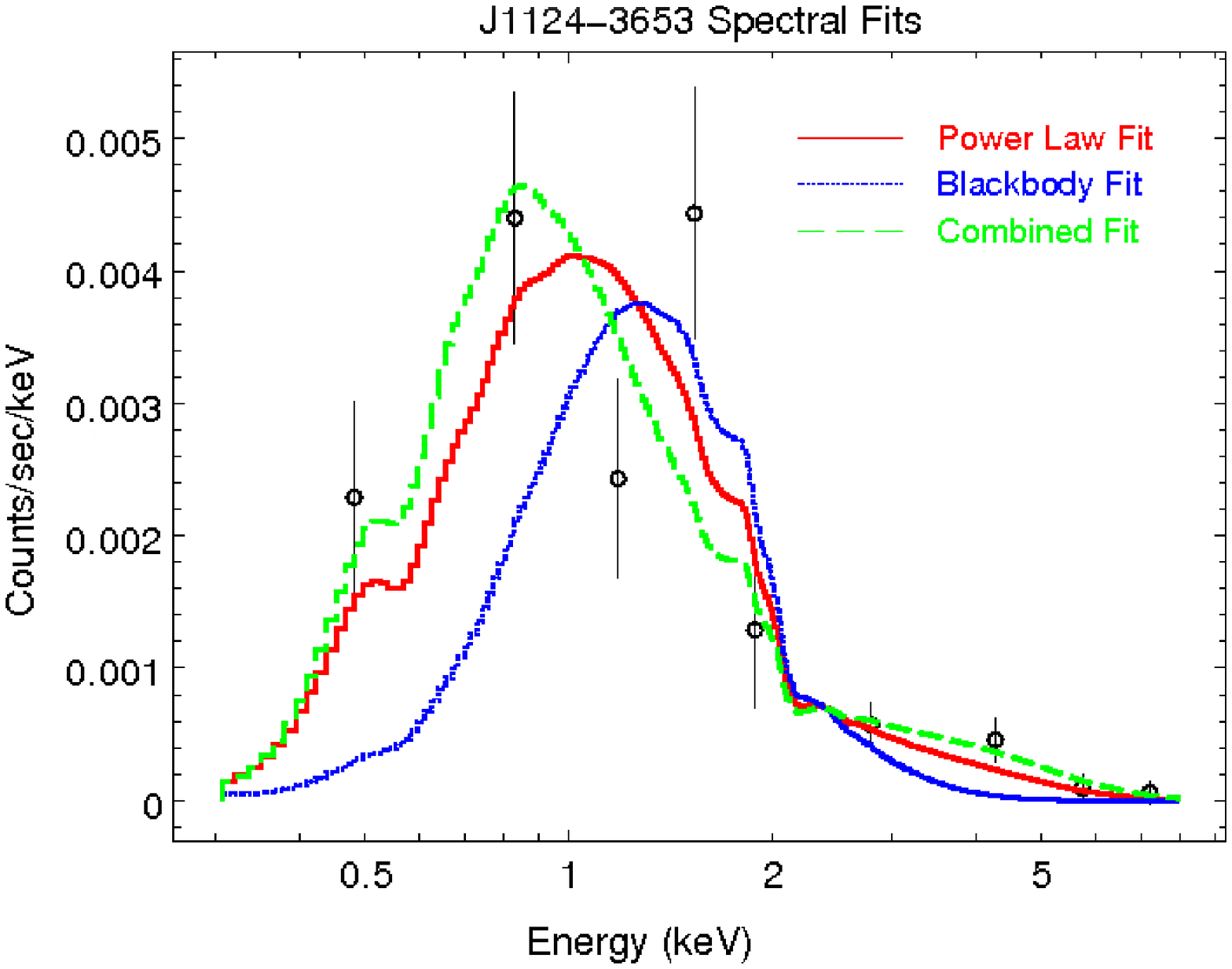}
\caption{Same as Fig.~1 but for PSR~J1124$-$3653.
%Count rate versus orbital phase and spectral fits for PSR~J1124$-$3653. We detected 140 counts and see significant, yet broad, orbital variation. Also shown are the radio eclipse start and stop times (dotted lines) at orbital phases of 0.16 and 0.37, respectively. The K-S test gives a probability of 0.10 of the lightcurve being drawn from a uniform distribution. Also shown are our fits for the spectrum. {\it The F-test shows that the power-law fit is the preferred fit}. Fit parameters are summarized in Table~\ref{table:table2}. {\it Non-Poissonian errors were applied to bins with less than 10 counts}. All error bars correspond to 1$\sigma$ errors.
%GENERAL: NON-P
\label{fig:fig3}}
\end{figure}

The lightcurve shows marginal orbital variability as evidenced by the K-S test, which yields a probability of 0.10 of being drawn from a uniform distribution. Although the lowest count rate occurs at an orbital phase of 0, there is a local minimum near an orbital phase of 0.25 (superior conjunction), which coincides with the radio eclipse phase (shown in Figure~\ref{fig:fig3}). Aside from these minima, the lightcurve is constant within the 1$\sigma$ error bars. A two-dimensional K-S test yields the probability of being a point source of 0.10 in the x-direction and 0.65 in the y-direction. Although the probability of the source being drawn from the same distribution as the PSF in the x-direction appears low, we note that the source is actually narrower than the PSF in the x-direction (consistent with Poisson variations) and therefore conclude that there is no evidence for extended emission for PSR~J1124-3653. The spectrum is well fit by a simple power law while a blackbody fit is formally unacceptable and results in a very high temperature. 
Since the count rate is higher than for PSR~J0023+0923, we fix temperature but let $\Gamma$ vary for the combined fit.  Most of the flux from the combined fit is assigned to the power-law component.

\subsection*{PSR~J1810+1744}\label{sub:J1810+1744}

\begin{figure}
\centering
\includegraphics[scale=.4]{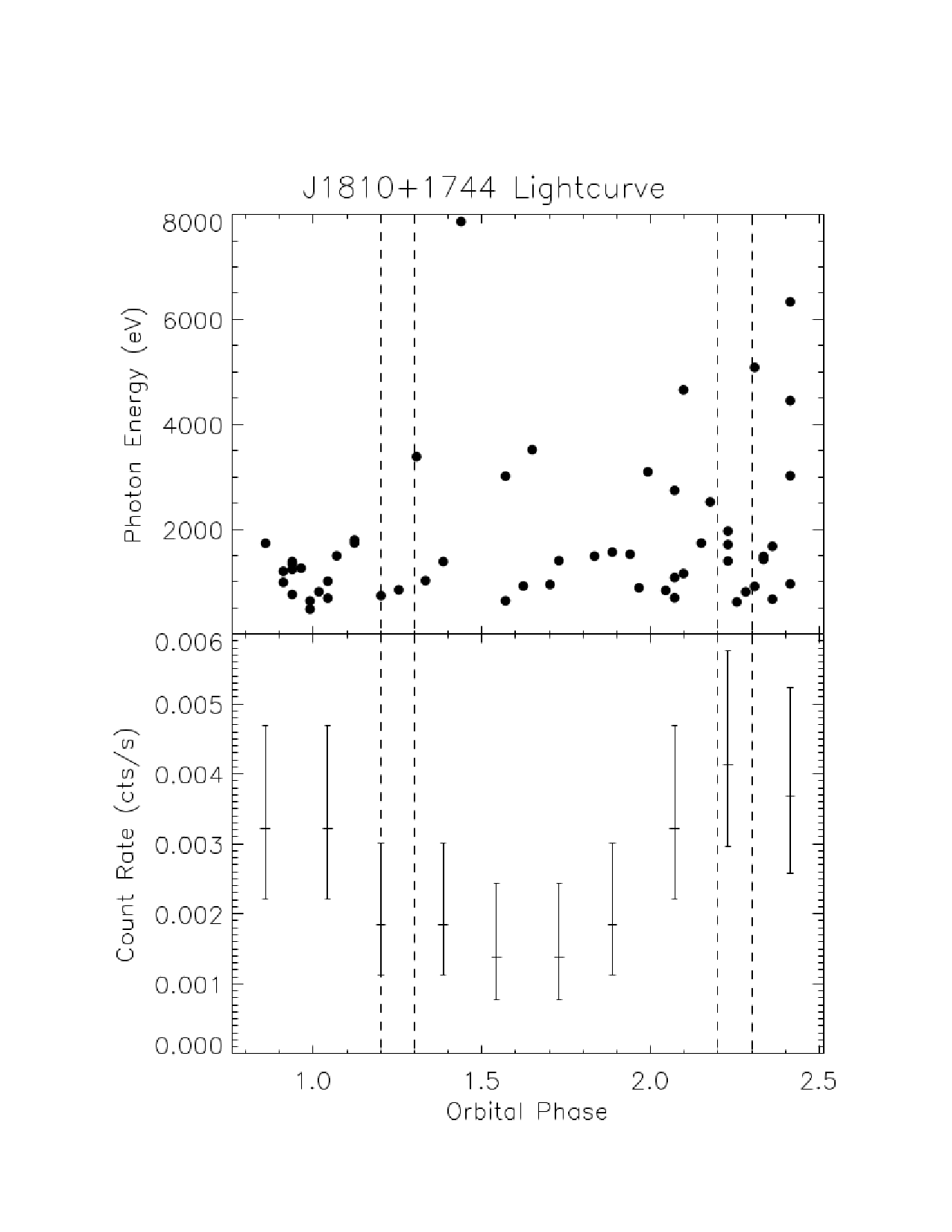}
\includegraphics[scale=.4]{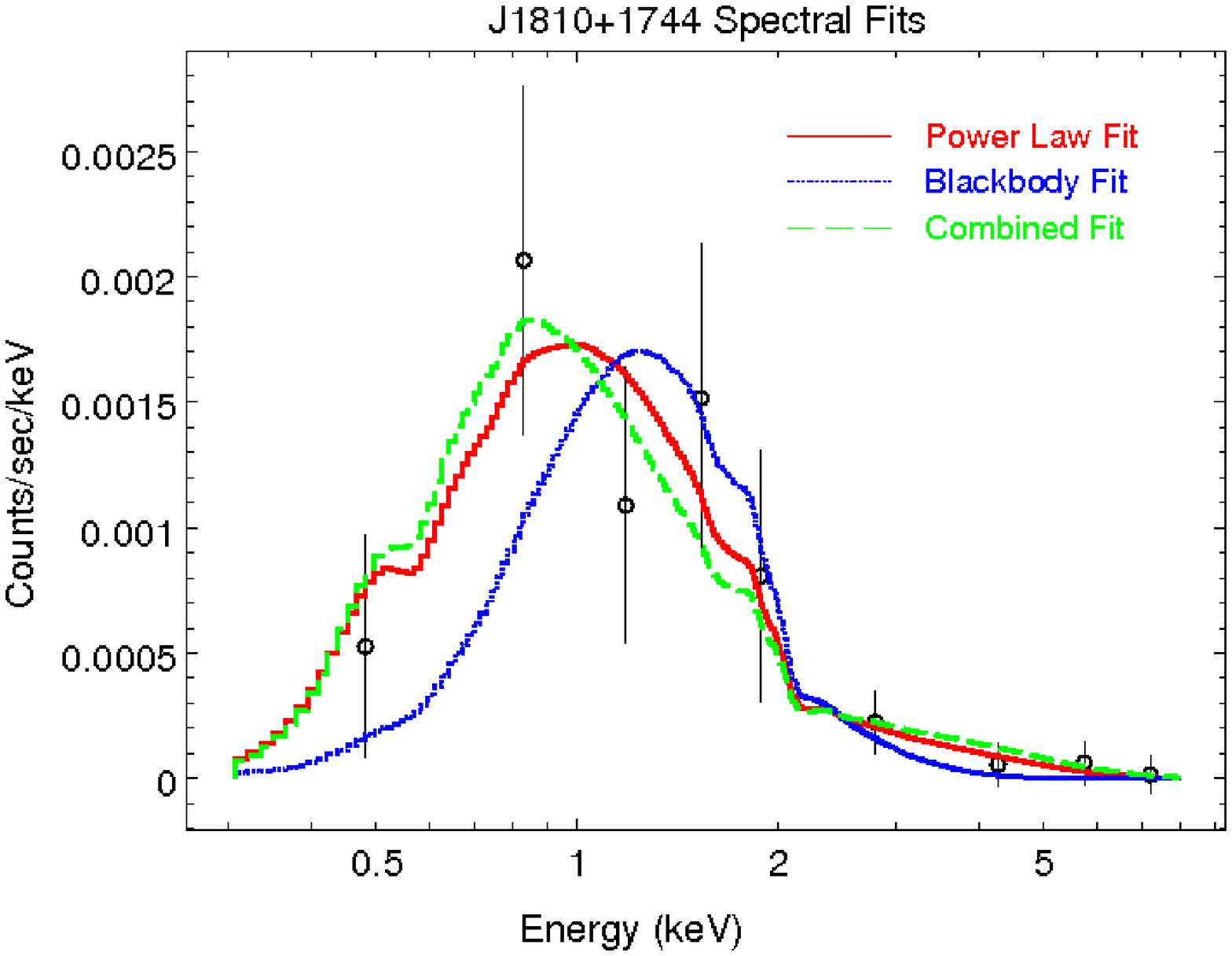}
\caption{Same as Fig.~1 but for PSR~J1810+1744.
%Count rate versus orbital phase and spectral fits for PSR~J1810+1744. We detected 55 counts  %and see significant, yet broad, orbital variation. Also shown are the radio eclipse start and stop %times (dotted lines), at orbital phases of 1.2 and 1.3, respectively. The K-S test gives a %probability of 0.08 of the lightcurve being drawn from a uniform distribution. Also shown are %our fits for the spectrum. The F-test shows that the power-law fit is the preferred fit. Fit %parameters summarized in Table~\ref{table:table2}. Non-Poissonian errors were applied to bins %with less than 10 counts in the bin. All error bars correspond to 1$\sigma$ errors.
%GENERAL: NON-P.
\label{fig:fig4}}
\end{figure}

The lightcurve does not look obviously uniform and the K-S test gives this lightcurve a probability of 0.43 of being drawn from a uniform distribution. The variation in the lightcurve is very broad, covering most of the orbit, making it unlikely that the orbital variation can be attributed to eclipsing of the intrabinary shock emission by the companion. The soft lightcurve (0.3 -- 2 keV) by itself does not show strong evidence for orbital variability. 
%TOPETE: K-S TEST RESULT? STRANGE THAT THE SOFT SHOWS MORE THAN COMBINED.
%Ah, you're right to be skeptical. The KS test says there's a probability of 0.45 of the soft lightcurve being drawn from a uniform distribution, which seems to indicate no significant orbital variability. When I wrote that, I was going by eye, which wasn't the best idea.
 The two-dimensional K-S test yields the probability of being a point source of 0.23 in both directions. Therefore, we conclude that there is no clear evidence for extended emission.
%We fit PSR~1810+1744's spectrum with a blackbody model, a power-law model, and a combined model with power-law and blackbody components. As with PSR J0023+0923, the low number of counts (55) resulted in an overdetermined fit, however in this case there were 5 counts observed above 4 keV, which is strong evidence for a power-law component, and the blackbody fit resulted in an unacceptably high temperature. 
We again fix temperature and $\Gamma$ for the combined fit, with most of the flux from the combined fit coming from the power-law component.

\subsection*{PSR~J2215+5135}\label{sub:J2215+5135}

\begin{figure}
\centering
\includegraphics[scale=.4]{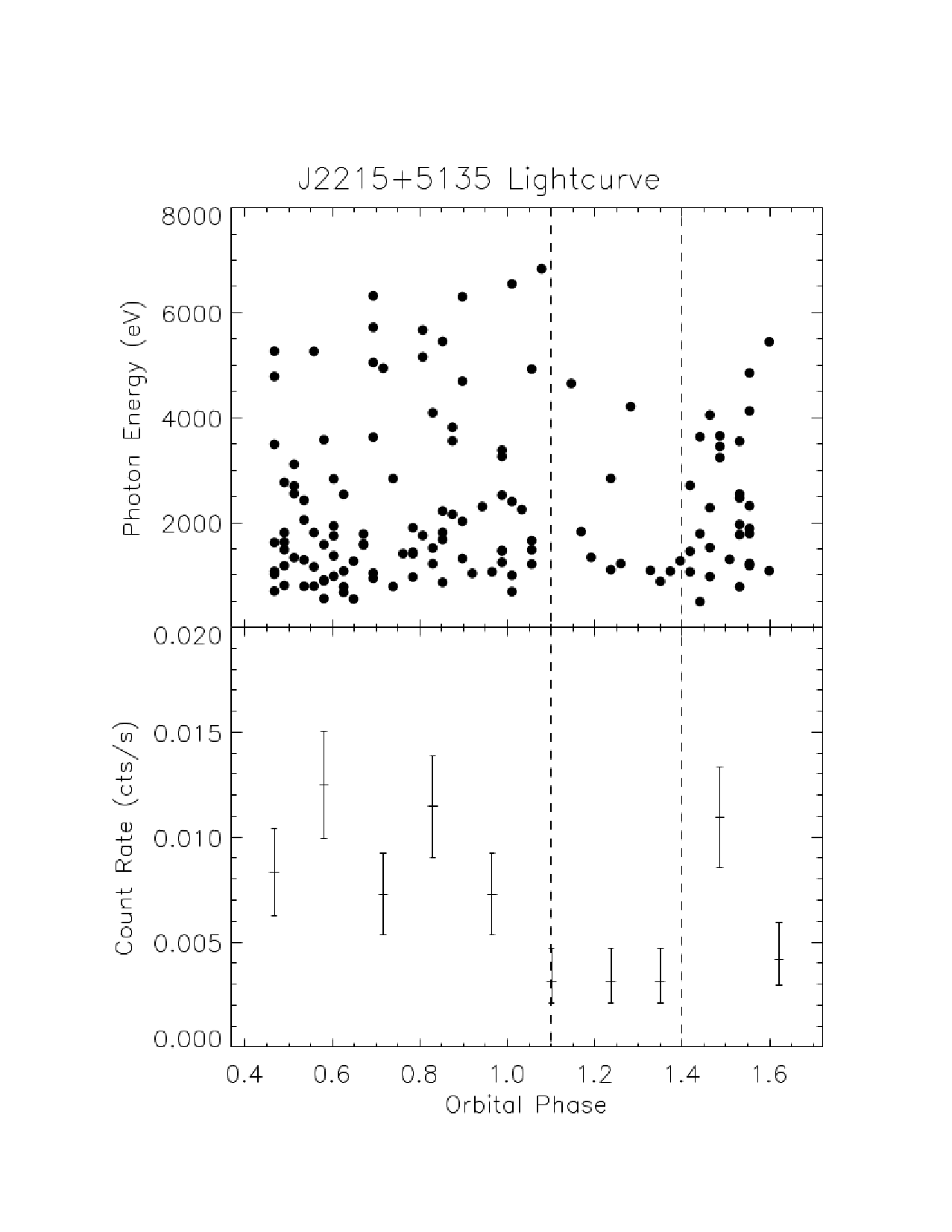}
\includegraphics[scale=.4]{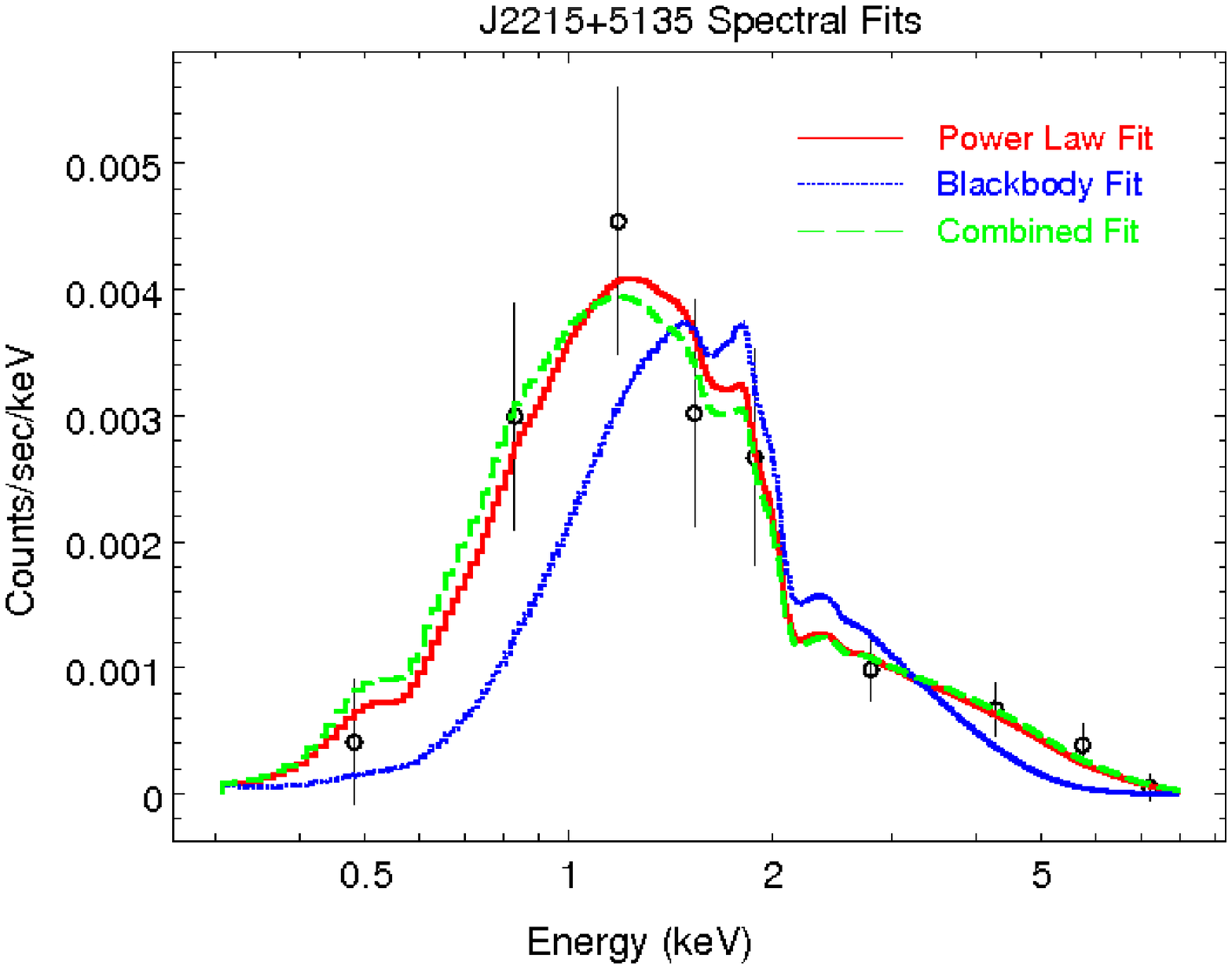}
\caption{Same as Fig.~1 but for PSR~J2215+5135.
%%Count rate versus orbital phase and spectral fits for PSR~J2215+5135. We detected 133 counts and see significant orbital variation during orbital phases from 1.1 to 1.35. Also shown are the radio eclipse start and stop times (dotted lines), at orbital phases of 1.1 and  1.4, respectively.  The K-S test gives a probability of $1\times{10^{-6}}$ of the lightcurve being drawn from a uniform distribution. Also shown are our fits for the spectrum, fit with a power-law model, blackbody model, and a combined model with power-law and blackbody components. {\it The F-test shows that the power-law fit is the preferred fit.} 
%Fit parameters are summarized in Table~\ref{table:table2}. Non-Poissonian errors were applied %to bins with less than 10 counts in the bin. All error bars correspond to 1$\sigma$ errors.
%GENERAL: NON-P.
\label{fig:fig5}}
\end{figure}

The single redback in our sample has a lightcurve which is clearly not uniform and the K-S test confirms this by yielding a probability of 0.04 of being drawn from a uniform distribution. Both the hard and soft lightcurves include clear minima at the same orbital phase as the radio eclipse.
A two-dimensional K-S test yields the probability of being a point source of 0.19 in the x-direction and 0.27 in the y-direction. Therefore, we conclude that there is no strong evidence for extended emission.
The spectrum is very hard, with a clear power-law tail. The blackbody fit resulted in a much higher $\chi^{2}$ value and an unacceptably high temperature. 
%As with the other sources, PSR~J2215+5135 was fit with a blackbody model, a power-law model, and a combined model with power-law and blackbody components.
We fix temperature, but let $\Gamma$ vary for the combined fit. The flux from the combined fit is again dominated by the power-law component.

\subsection*{PSR~J2256$-$1024}\label{sub:J2256-1024}

\begin{figure}
\centering
\includegraphics[scale=.4]{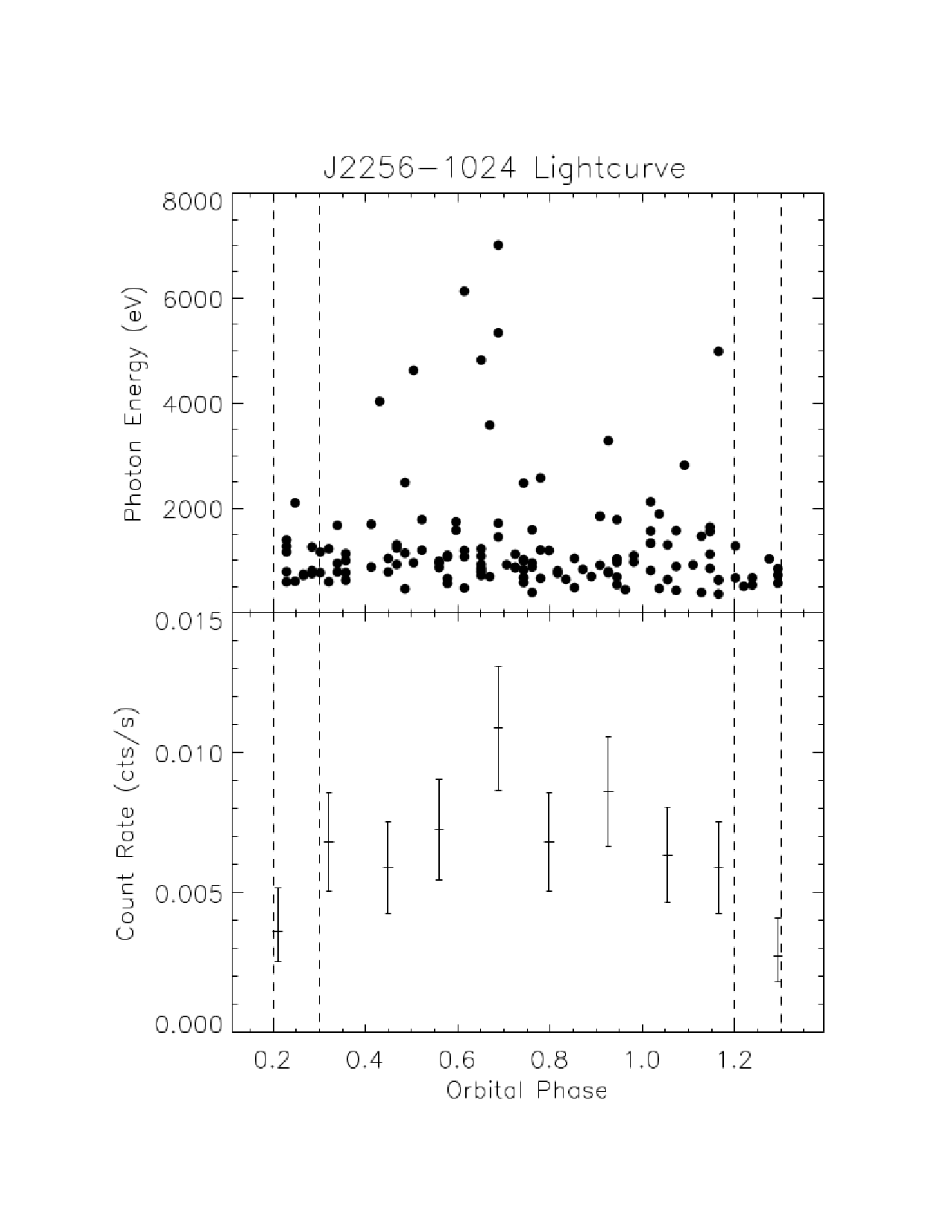}
\includegraphics[scale=.4]{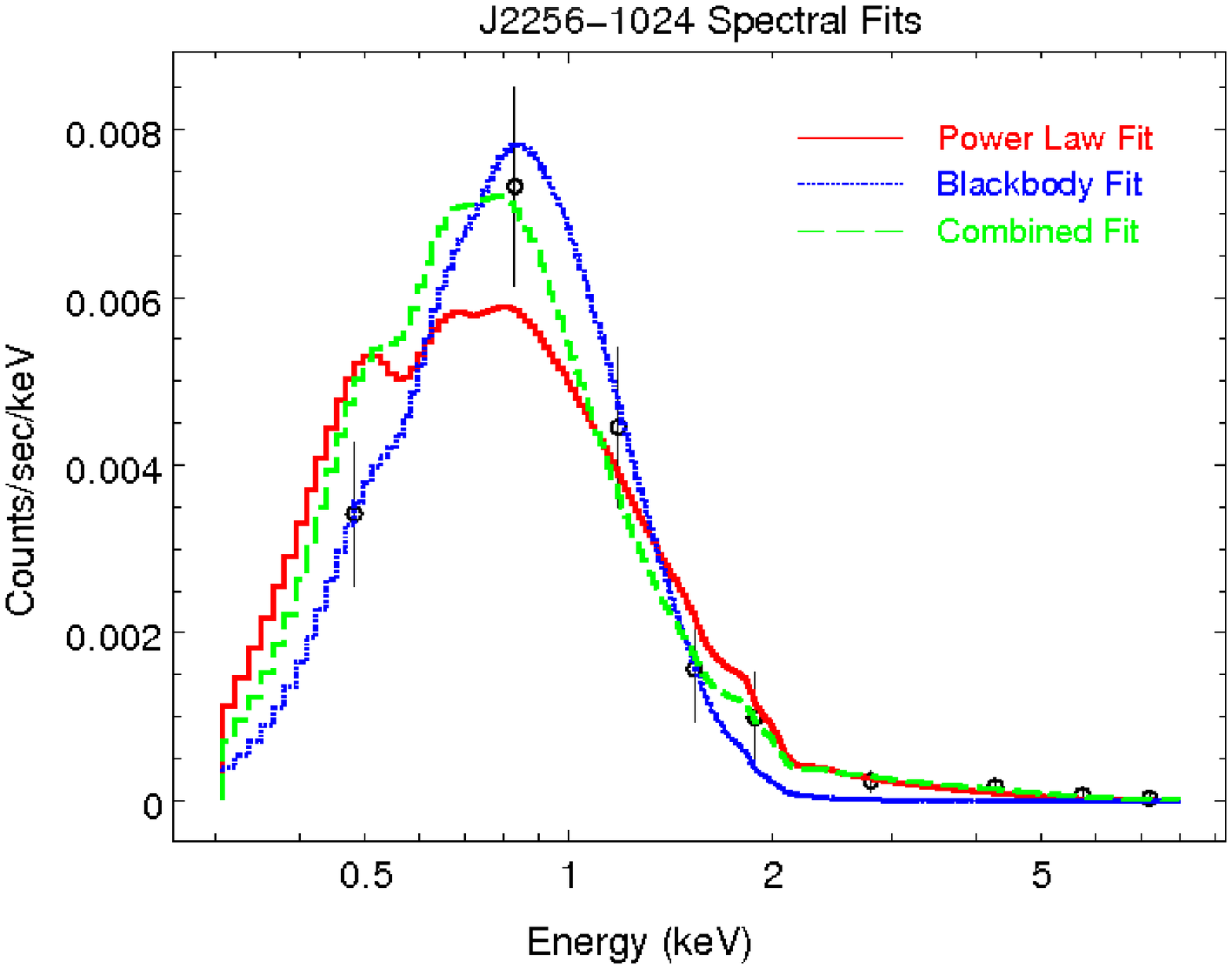}
\caption{Same as Fig.~1 but for PSR J2256$-$1024.
%%Count rate versus orbital phase and spectral fits for PSR J2256$-$1024. We detected 133 counts and see significant dips in the lightcurve at orbital phases of 0.2 and 1.3. Also shown are the radio eclipse start and stop times (dotted lines) at orbital phases of 0.2 and 0.3.  The K-S test gives a probability of $8.8\times{10^{-3}}$ of the lightcurve being drawn from a uniform distribution.  Also shown are our fits for the spectrum, which was fit with a power-law model, a blackbody model, and a two-component power-law and blackbody model. The F-test shows that the combined model is the preferred model. {\it Non-Poissonian errors were applied to bins with less than 10 counts in the bin.} All error bars correspond to 1$\sigma$ errors.
%GENERAL: NON-P.
\label{fig:fig6}}
\end{figure}

The lightcurve has clear minima near orbital phases of 0.25 and 1.25 and the K-S test gives this lightcurve a probability of $8.8\times{10^{-3}}$ of being drawn from a uniform distribution. Although the dip around 0.25 is pronounced, we only have a single coverage of the minimum. Although we do not see the same dip in the soft lightcurve, the hard lightcurve does seem to have dips at the same orbital phases that the general lightcurve has.  The dips coincide with the measured radio eclipses. A two-dimensional K-S test yields the probability of  being a point source of 0.96 in the x-direction and 0.60 in the y-direction. Therefore, we conclude that there is no evidence for extended emission.
%PSR~J2256$-$1024 was also fit with a blackbody model, a power-law model, and a combined model with power-law and blackbody components. 
Both the power-law and blackbody fits are acceptable, with a reasonable temperature and a somewhat steep spectral index. However, around 5\% of the photons are above 4~keV, which, along with the orbital variability, suggests a significant power-law spectral component. 
We fix temperature, but let $\Gamma$ vary for the combined fit. The F-test prefers the combined fit over the power-law fit with a significance of 0.95, and the flux from the combined fit is fairly evenly split between blackbody and power-law components.

\begin{flushleft}
\begin{figure}
\includegraphics[scale=.5]{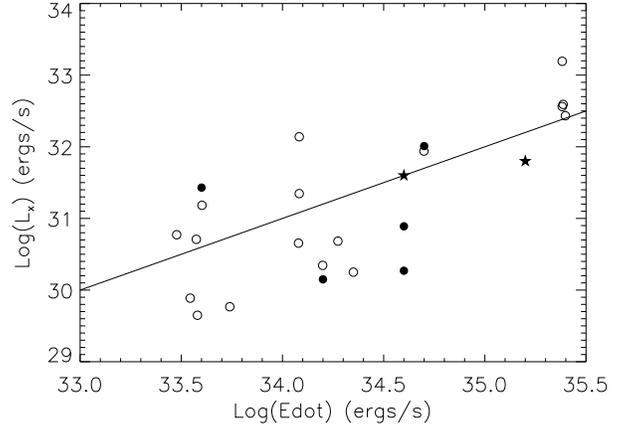}
\caption{Luminosity versus spin-down energy loss rate $\dot E$ for 24 X-ray detected MSPs. PSRs~J0023+0923, J1124$-$3653, J1810+1744, J2215+5135, and J2256$-$1024 are shown as filled circles. Open circles represent all other Galactic MSPs from which X-ray emission has been detected. Spin-down energy loss rates were calculated using period derivatives that have been corrected for proper motion, and luminosities were calculated using distances derived from parallax measurements, where possible. PSRs~J1023+0038 and B1957+20 are shown as stars for comparison. Luminosities of all sources were converted to equivalent luminosities in the 0.3 -- 8.0 keV range using WebPIMMS. The line represents 0.1\% efficiency.}
\label{fig:fig7}
\end{figure}
\end{flushleft}

\section{Discussion and Conclusions}\label{sec:conclusions}
X-ray emission has been detected from roughly 50 MSPs\footnote{See http://astro.phys.wvu.edu/XrayMSPs for a full list of sources and for parameters used to calculate the luminosities in Figure 6.}. The emission can be described by either blackbody or power-law models and can originate from the neutron star surface (in the case of a blackbody model) or from the magnetosphere or an intrabinary shock (in the case of a power-law model). 
%{\it In the case of magnetospheric emission from the NS surface, we expect power-law indices %of $\sim$ 2, while power-law indices for intrabinary shock emission are often lower, around %$1$ to $1.5$ (\cite{bgv05,bah+11}).}
%%I'M NOT SURE THIS IS TRUE. FOR ONE THING, SEVERAL PULSARS HAVE PULSED SPECTRA  \ll 2, AND FOR ANOTHER, WE ONLY HAVE A COUPLE OF EXAMPLES OF INTRABINARY SHOCK EMISSION.  
 We expect emission from the neutron star's surface and magnetosphere to be steady on timescales longer than the pulse period, and expect orbital modulation in the case of emission from an intrabinary shock. This modulation can be due to Doppler boosting of the flow within the shock, synchrotron beaming, or obscuration by the companion. In the first two cases, we would expect enhanced emission when the flow is coming towards us. Since there is only a weak outflow from the companion, we would expect a Mach cone pointed away from the pulsar with its head near the point on the companion star closest to the pulsar. For a nearly Roche-lobe filling companion, this would be near the L1 point. We might therefore expect a minimum near inferior conjunction (orbital phase 0.75), and, depending on inclination, a broad peak roughly centered around superior conjunction (orbital phase 0.25). However, the orbital motion would cause the Mach cone to be swept back, in which case a broad enhancement after superior conjunction may result. The companion could also obscure part of the shock near superior conjunction, causing an X-ray dip. The duration and depth of the dip would depend on the ratio of the companion radius to the intrabinary separation, as well as the inclination angle of the system.   With any of the above mechanisms, we would expect little if any change in the observed spectrum of the shock. 
%%wis primarily of hard photons (as seen in \cite{2008AIPC..983...64B}). Typically, this %modulation is manifested as a dip in the X-ray emission
%around an orbital phase of 0.25, when the pulsar is behind the companion star (see %\cite{2011ApJ...742...97B} and \cite{2005ApJ...630.1029B}).
Extended X-ray emission due to the pulsar wind shocking the interstellar medium has been detected from some MSP binary systems, and this would also be expected to be steady.

For emission arising from an intrabinary shock, the angle subtended on the pulsar's sky by the  companion determines the fraction of the wind involved in the shock as well as affecting the X-ray light curve. If the companion is Roche-lobe filling, this fraction depends only on the masses of the binary components, which can be inferred from the timing modulo the inclination angle of the system. Modeling of the optical lightcurve of the companion can constrain both the inclination angle and the Roche-lobe filling factor of the companion. \citet{bvr+13}
 have made optical studies of all of our sources except for PSR~J1124$-$3653, and compare them to PSR~J1023+0038 and PSR~B1957+20. All except PSR~J0023+0923 and possibly PSR~J2256$-$1024 seem to be nearly filling their Roche-lobe. The radius of the companion to PSR~J0023+0923 may well be less than 2/3 its Roche-lobe radius, and the diameter subtends only $\sim$8$^{\circ}$. PSR~J2256$-$1024 subtends $\sim$11$^{\circ}$, PSR~J1810+1744 $\sim$15$^{\circ}$, and PSR~J2215+5135 $\sim$26$^{\circ}$. All are viewed at moderate inclination angles $i\sim$45$^{\circ} - 70^{\circ}$. Although we do not have optical information on PSR~J1124$-$3653, if it is nearly Roche-lobe filling as well, it would subtend $\sim$15$^{\circ}$.  

 We detect X-ray emission from all five observed MSP binary systems 
 (PSRs~J0023+0923, J1124$-$3653, J1810+1744, J2215+5135, and J2256$-$1024). None of the pulsars show strong evidence for extended emission.
In most cases, there is strong evidence for non-thermal emission, with power-law indices $\sim$$1-2$ consistent with intrabinary shock emission, similar to what is seen in the modulated emission from PSR B1957+20 and PSR J1023+0038  (\cite{bgv05}, \cite{bah+11}). While not well constrained given our low statistics, the ratio of non-thermal to thermal flux from our sample seems to roughly scale with the solid angle subtended by the companion. We also note that the X-ray luminosities for our sources are comparable to other pulsars with similar spin-down energy loss rates (see Figure~\ref{fig:fig7} and \citealt{pkgw07}). 

Two of the five pulsars show strong evidence for orbital modulation. PSR J2215+5135 shows an X-ray dip for roughly a quarter of the orbit around the radio eclipse. This is seen in both hard and soft lightcurves. Given the large angle the companion subtends on the pulsar sky, we should expect comparatively more intrabinary shock emission and a broader X-ray dip than the other sources, as well as even longer radio eclipses, even at high frequencies. Observations at 2 GHz with the Green Bank Telescope show it to be eclipsed for roughly 1/3 of the orbit %\citet{h+13}
.     Similarly, we see a dip in the X-ray lightcurve around the radio eclipse for PSR~J2256$-$1024. This dip is more pronounced in the hard lightcurve. We therefore conclude that the power-law spectral components for these two pulsars are primarily due to intrabinary shock emission. Another two of the five pulsars show marginal evidence for orbital variability, with broadly sinusoidal lightcurves. For PSR J1124$-$3653, the emission appears to peak around half an orbit after the radio eclipse, but comparing the beginning of the observation to the end also hints at orbit-to-orbit variability.
% One dip anomalously occurs at an orbital phase of 0.0. A second dip occurs at an orbital phase %of 0.25, which is aligned with the radio eclipse. 
This could be due to intrabinary shock emission, though a longer observation is necessary to further probe this. PSR~J1810+1744 shows broad orbital variability around the orbit, with possible orbit-to-orbit variations. Given the often chaotic nature of wind shocks, this is only to be expected, as has been observed in both PSR B1957+20 and PSR J1023+0038. 

The lightcurve of PSR~J0023+0923 is nearly uniform, although due to the small number of counts, it is hard to make any concrete conclusion about variability. However, it also shows no evidence for radio eclipses and no evidence for emission above 2.5~keV. Given the companion's small angular extent and apparent under-filling of its Roche-lobe, meaning the surface material is much more strongly gravitationally bound than for the Roche-lobe filling systems, any contribution from shock emission is expected to be small.  

We conclude that the emission from both PSRs~J1124$-$3653 and J2256$-$1024 is likely due to a combination of thermal emission from the neutron star surface and power-law emission from an intrabinary shock. The emission from PSR~J2215+5135 is consistent with being due primarily to an intrabinary shock. The temperatures and power-law indices derived are consistent with previous fits to neutron star spectra.  For all three of these pulsars, a small magnetospheric contribution is also possible. Further X-ray observations with better timing resolution are necessary to determine this. Given the small number of counts for PSRs~J0023+0923 and J1810+1744, it is difficult to make conclusions on the origin of the X-ray emission. However, given the emission from PSR~J0023+0923 seems likely to be predominantly thermal, it is likely we are only seeing emission from the pulsar itself, with essentially no contribution from a shock.

The small number of photons detected from all of these sources prohibits a more detailed study or detailed geometrical modeling. However, in all cases the emission is dominated by an unresolved source, and likely comes from within the system with little or no contribution from an extended wind nebula. Therefore, future studies covering multiple orbits with any of the current imaging X-ray telescopes are highly desirable. 
%X-ray observations of black-widow systems are rare, with only PSR~B1957+12 studied in detail. %Further observations of these five systems will allow us to model the orbital variability in detail, %determine precise spectra, constrain orbital spectral dependence, and perform a deep search %for extended emission below the level seen for PSR~B1957+20. They will also allow us to %determine how the currently poorly constrained X-ray flux depends on $\dot{E}$, M$_{\rm %c,min}$, and the pulsar magnetic field. They will also give us insight into the shock mechanism, %allow us to search for faint extended emission, determine the contributions to the  flux from %magnetospheric, intrabinary shock, and thermal emission, and determine whether the unusual %pulsar PSR~J2215+5135 is a black widow in the early stages of companion ablation or a newly %born MSP in the late stages of recycling by comparing its properties to those of other black %widows and newly formed MSPs (such as PSR~J1023+0038).
%Optical observations of these sources
%are necessary to determine companion masses and hence inclinations, essential for future %modeling.

Support for this work was provided by the National Aeronautics and Space Administration through Chandra Award Number GO1-12061A and GO2-13056X issued by the Chandra X-ray Observatory Center, which is operated by the Smithsonian Astrophysical Observatory for and on behalf of the National Aeronautics Space Administration under contract NAS8-03060. This research has made use of NASA's Astrophysics Data System. This research has made use of data obtained from the High Energy Astrophysics Science Archive Research Center (HEASARC), provided by NASA's Goddard Space Flight Center. JWTH acknowledges funding through the Netherlands Foundation for Scientific Research (NWO). MAM gratefully acknowledges support from Oxford Astrophysics while on sabbatical leave.

\bibliography{journals,psrrefs,modrefs}
\bibliographystyle{apj}

\end{document}